\documentstyle[preprint,revtex]{aps}

\begin{document}

\begin{title}
Propagation of a hole on a N\'eel background
\end{title}

\author{E. M\"uller-Hartmann  and  C. I. Ventura~$^*$}
\begin{instit}
Inst. f\"ur Theoretische Physik, Univ. zu K\"oln,
Z\"ulpicher Stra\ss e 77, K\"oln 50937, Germany.
\end{instit}

\begin{abstract}
We analyze the motion of a single hole on a N\'eel background, neglecting
spin fluctuations. Brinkman and Rice studied this problem on a cubic
lattice, introducing the retraceable-path approximation for the hole
Green's function, exact in a one-dimensional lattice. Metzner et al.
showed that the approximation also becomes exact
in the infinite-dimensional limit.
We introduce a new approach to this problem by resumming the Nagaoka
expansion of the propagator in terms of non-retraceable
skeleton-paths dressed by
retraceable-path insertions. This resummation
opens the way to an almost quantitative
solution of the problem in all dimensions and, in particular sheds
new light on the question of the position of the band-edges.
We studied the motion of the hole on a double chain and
a square lattice, for which deviations from the retraceable-path
approximation are expected to be most pronounced.
The
density of states is mostly adequately accounted for by the
retra\-ce\-able-path approximation. Our band-edge determination
points towards an absence of band tails extending to the Nagaoka energy
in the spectrums of the double chain and the square lattice.
We also evaluated the spectral density and the
self-energy, exhibiting k-dependence due to finite dimensionality.
We find good agreement with recent numerical results obtained
by Sorella et al. with the Lanczos spectra decoding method. The method
we employ enables us to identify the hole paths which are
responsible for the various features present in the density of states
and the spectral density.
\end{abstract}

\narrowtext

\section{INTRODUCTION}
\label{sec:intro}

The discovery of high-temperature superconductivity has
intensified the interest in the study of strongly correlated
electron systems. Among the important related subjects awaiting solution
is the dynamics of holes in the presence of a spin background.
Among pioneering works on the problem one can cite that by
Bulaevskii, Nagaev and Khomskii \cite{nagaev} who, studying the motion of
a hole on an antiferromagnetic background, with the neglect of quantum
fluctuations, introduced the ``string potential" concept.
As the hole moves on the N\'eel background it leaves behind
a string of overturned spins, thus
increasing the exchange energy proportionally to the
length of the string. This can
lead to self-trapped (or, as they were called in Ref.~\cite{nagaev},
``quasi-oscillator")
states centered at the original position of
the hole. Two years later, Brinkman
and Rice \cite{brice} studied the motion of a single hole on different
spin backgrounds,
employing the Nagaoka expansion for the
hole propagator \cite{nagaoka}. This involves
contributions by hole paths restoring the spin
background, classified according
to their length. For a cubic lattice and ferro-, antiferromagnetic and random
spin backgrounds,
they calculated the exact
contributions of hole paths up to length 10, neglecting spin fluctuations.
The density of states obtained was compared with the one resulting
from considering only the self-retracing paths of the hole,
the ``retraceable-path approximation"(denoted RPA throughout this paper).
In this
approximation, which does not distinguish between different spin backgrounds,
one can exactly sum the series for the propagator, which involves
only local contributions, thus obtaining no
momentum(k)-dependence: the motion is completely incoherent.
With the RPA Brinkman and Rice \cite{brice} were
able to
account for the body of the band in the case of a N\'eel
background. Less agreement was obtained \cite{brice}
for the random background, while
the ferromagnetic case reduces to the uncorrelated single particle
motion in a tight-binding band which generally is quite different
from the RPA description.
 They also observed that the RPA became exact
in the one-dimensional case where, for any spin background,
the free-particle result is obtained. The band edges appear at the
so called Nagaoka energies.

In a series of recent works \cite{mvollh,svollh} the motion of a
hole on a spin background was analyzed in the case of
infinite dimensions. In particular, Metzner et al. \cite{mvollh}
estimated the contributions to the propagator in terms of the
dimension(D). Neglecting spin fluctuations, the only non-retraceable
contributions
which would remain finite at D=$\infty$
would be those of loop paths
circulated only once. These would certainly not preserve a N\'eel ordered
spin background, so that the RPA becomes the exact solution
of the problem in this case. For the N\'eel background at finite dimensions,
the lowest order corrections to the RPA density of states in the case of a
hypercubic lattice would come
from circulating elementary square plaquettes three times (thus
of order $\sim 1/D^{4}$).
More important are the non local corrections to the propagator,
the lowest of which is of order $\sim 1/D$ (the hole can propagate to
nearest neighbour sites on the same sublattice circulating
an elementary square plaquette one and a half times). The latter
corrections would give rise to a k-dependent self-energy for the hole.
Trugman \cite{trugman}
already mentioned these non-retraceable paths in his study of the
two-dimensional Hubbard antiferromagnet. Quite different physics is expected
when the quantum fluctuations associated with transverse exchange
interactions are taken into account from the start. This problem
has been addressed in many recent works, among others see
Refs.~\cite{siggia,ruck,kane,tsunetsugu,maekawa,horsch,gagliano}.

Taking into account the results mentioned above for hole motion on a N\'eel
background in the absence of quantum fluctuations,
 we propose here a new method to evaluate the local and non-local parts
of the propagator in order to analyze the departures from the RPA for reduced
dimensionality. In Section~\ref{sec:method} we describe the method employed
to obtain these quantities. Basically, the
Nagaoka expansion of the propagator is rewritten in terms of
 non-retraceable skeleton-paths dressed by retraceable paths.
Though one can explicitly evaluate the contribution
of a dressed skeleton diagram of given length,
one has to determine numerically the
numbers of bare background-restoring skeleton-paths.
The results obtained with this method for the double chain
and the square lattice are presented in
Section~\ref{sec:results}. We can quite reliably determine
the band edges, and our results place them at an intermediate value
between the RPA edge and the Nagaoka energy.
In general,
we find good agreement with the recent numerical
results of Sorella et al. \cite{sorella}, obtained with the Lanczos
spectra decoding method.
The method we employ enables us to determine the
origin of the distinctive features exhibited by the density of states
and the spectral density.
Thus, we can identify the
relevant paths responsible for the k-dependence at reduced dimensionality.
In Section~\ref{sec:summary} we summarize our results.

\section{ANALYTICAL METHOD}
\label{sec:method}

In this section we describe the method employed to evaluate
the hole propagator on a N\'eel background, in which spin
fluctuations are neglected.

The Nagaoka expansion \cite{nagaoka,mvollh} for the Green's function $G$
describing the propagation of a hole inserted into an arbitrary fixed spin
configuration $\mid s >$, takes the following form:

\begin{eqnarray}
G_{i,j}^{s}(w) \, & \, = \, & \, \sum_{\sigma} <  s \mid c^{+}_{i, \sigma}
\frac{1}{w - H} c_{j, \sigma} \mid s >
\nonumber \\
\, & \, = \, & \, \frac{1}{w}\left[ \delta_{i,j} + \sum_{n=1}^{\infty}
A_{i,j,n}^{s} \left(
\frac{-t}{w} \right)^{n} \right] , \nonumber \\
H \, & \, = \, & \, - t \sum_{<i,j>,\sigma} [ \,
(1 - n_{i, -\sigma})
c^{+}_{i, \sigma}
c_{j, \sigma} (1 - n_{j, -\sigma}) \nonumber \\
\, & \, & \, + \, h.c. \, ] .
\label{ges}
\end{eqnarray}

Above, $c^{+}_{i, \sigma} $ ($c_{i, \sigma}$) creates (annihilates)
a particle with spin $\sigma$ at the lattice
site $i$, while $H$ is the Hamiltonian describing the
correlated hopping of a particle
between nearest neighbour sites.
The coefficients $A_{i,j,n}^{s}$
denote the number of distinct $n$-step paths of the hole in the
spin background $s$ which start at site $j$ and end
at site $i$, restoring the original spin configuration after the last
step. In particular we will consider the case of a N\'eel spin
background, $s = N$, where propagation is only possible between sites
belonging to the same sublattice.

As we mentioned in the previous section, the paths in which the hole
completely retraces its steps are the only relevant ones at $D = \infty$
on a N\'eel background \cite{mvollh}. Even at finite dimensions they
would determine the leading terms of the local Green's function.
Following Brinkman
and Rice \cite{brice} one can derive the RPA propagator
 by introducing the following
irreducible ``self-energy" $S$ (here irreducibility means that
the path cannot be split up into two or more consecutive retraceable paths):
\begin{eqnarray}
G_{RPA}(w) \, & \, = \, & \, \frac{1}{w (1 - S(w))} .
\label{rpadef}
\end{eqnarray}
The lowest order RPA self-energy will derive from
 the contributions of single jumps to
a nearest neighbour and immediate return to the origin, so that:
\begin{eqnarray}
S^{0}(w) \, & \, = \, & \, Z (\frac{t}{w})^2 ,
\label{rpa0}
\end{eqnarray}
where $Z$ is the number of nearest neighbours.
The next order irreducible self-energy contribution consists of one jump
to a nearest neighbour
of the origin followed by a further jump in any direction different
from the previous jump (thus having $(Z-1)$ possibilities for this step)
before retracing the whole path.
In general, one can write the RPA irreducible
self-energy as:
\begin{eqnarray}
S(w) \, & \, = \, & \, Z (\frac{t}{w})^2 C(w) , \label{ssrpa1} \\
C(w) \, & \, = \, & \, 1 + (\frac{Z-1}{Z}) S(w)
+ (\frac{Z-1}{Z})^2 S(w)^2 + ... \nonumber \\
\, & \, = \, & \, \frac{1}{1 - \left( \frac{Z-1}{Z} \right) S(w)} ,
\label{ssrpa2}
\end{eqnarray}
that is, it is written self-consistently in terms of the jump
to a nearest neighbour of the origin
dressed by retraceable paths, where $C$ is the RPA-dressed irreducible vertex
part.

The solution of the self-consistent system of equations (\ref{ssrpa1}) and
(\ref{ssrpa2}),
is \cite{brice}:
\begin{eqnarray}
S(w) \, & \, = \, & \, \frac{Z}{2 (Z-1)} \left[
1 - \sqrt{1 - 4 (Z-1)(\frac{t}{w})^2}
\right] .
\label{srpa}
\end{eqnarray}
Therefore the RPA propagator is given by:
\begin{eqnarray}
G_{RPA}(w) \, & \, = \, & \,
\frac{1}{w  - \frac{Z}{2(Z-1)} \left[ w - \sqrt{w^{2} -
4 (Z-1) t^{2}} \right] } . \nonumber \\
\,& \,&
\label{grpa}
\end{eqnarray}

 From this expression one can easily determine the band edges, $\pm w_{RPA}$,
of the energy spectrum obtained in the RPA for a hole hopping
on a N\'eel background as:
\begin{eqnarray}
w_{RPA} \, & \, = \, & \, 2  \sqrt{Z-1} t ,
\label{wrpa}
\end{eqnarray}
which exhibits a reduction from the free-particle value (Nagaoka energy):
$w_{FP}  =   Z t$ for $Z > 2$,
while the RPA coincides with the free-particle result in the
 one-dimensional ($Z = 2$) case, as
already mentioned in the previous section.

Having introduced
the retraceable-path approximation (RPA)
which gives a good approach to the local pro\-pa\-ga\-tor, we will
now describe the method we employed to obtain the Green's function
 of a hole moving on a
N\'eel background. In the Nagaoka expansion (\ref{ges}) we take into account
all background-restoring paths in the following way:
for each path-length, we take all possible non-retraceable
``skeleton"-paths which restore the N\'eel background,
and dress them by inserting all possible retraceable paths.
The RPA paths result from dressing the trivial skeleton-path of length $0$.
For all other skeleton-paths two types
of dressing have to be taken into account: 1) additional steps
within the skeleton-path (internal dressing), 2) irreducible
RPA vertex parts inserted at all vertices of the internally dressed
skeleton-path. To avoid double-counting one has to
distinguish between external and internal vertices in the
internally dressed skeleton
diagram. External vertices are connected by one link
to the skeleton which implies only one forbidden first step for all
retraceable-path insertions. The dressed ``external" vertex $C_{ex}(w)$
 is therefore correctly described by Eq.~(\ref{ssrpa2}).
 Internal vertices, however, are doubly connected
to the skeleton-path, such that two first steps are forbidden
for insertions
of the retraceable-path vertex dressings.
Therefore, the RPA-dressed irreducible internal vertex part is:
\begin{eqnarray}
C_{in}(w)
\, & \, = \, & \, \frac{1}{1 - \left( \frac{Z-2}{Z} \right) S(w)} .
\label{cint}
\end{eqnarray}

In order to perform the dressing we have to count the number of internal
dressings which can be considered for a skeleton
diagram of length $l$. For this we define
$N_{l}^{k}(m,n)$ as the number of paths of $m$ steps on a skeleton diagram
of length $l$, starting at site $0$, reaching site $k \, (0\leq k\leq l)$
after step $m$, and visiting the end-points $0$ and $l$ exactly $n$ times
(including the start at site $0$). In terms of these numbers the dressed
contribution of  a single skeleton diagram of length $l$ to the propagator
is given by
\begin{eqnarray}
G_{l}(w) \, & \, = \, & \, \frac{1}{w}\left[ \sum_{m,n} N_{l}^{l}(m,n)
(\frac{t}{w})^{m}
 C_{in}^{m+1-n}  C_{ex}^{n} \right] .
\label{gl}
\end{eqnarray}

For the numbers $N_{l}^{k}(m,n)$ the following recursive system of
equations holds:
\begin{eqnarray}
N_{l}^{0}(0,n) \, & \, = \, & \, \delta_{n,1}  ,\nonumber \\
N_{l}^{0}(m+1,n+1) \, & \, = \, & \, N_{l}^{1}(m,n) ,
\nonumber \\
N_{l}^{k}(m+1,n) \, & \, = \, & \, N_{l}^{k+1}(m,n) +
N_{l}^{k-1}(m,n) , \, \nonumber \\ & \, \, & \,
\qquad \qquad \qquad \qquad (0<k<l)
\nonumber \\
N_{l}^{l}(m+1,n+1) \, & \, = \, & \, N_{l}^{l-1}(m,n) .
\label{niter}
\end{eqnarray}

One can solve these equations by introducing the generating functions
\begin{eqnarray}
g_{k}(x,y) \, & \, = \, & \,  \sum_{m,n} N_{l}^{k}(m,n)
x^{m}  y^{n},
\label{genf}
\end{eqnarray}
and rewriting the equations (\ref{niter}) in terms of them. One obtains
\begin{eqnarray}
g_{0}(x,y) \, & \, = \, & \, y + x y g_{1}(x,y) , \label{g1} \\
g_{k}(x,y) \, & \, = \, & \, x \left( g_{k-1}(x,y) + g_{k+1}(x,y) \right) ,
\, \nonumber \\ & \, \, & \, \qquad \qquad
\qquad \qquad (0<k<l),   \label{g2} \\
g_{l}(x,y) \, & \, = \, & \, x y g_{l-1}(x,y) . \label{g3}
\end{eqnarray}

The general
solution of Eq.~(\ref{g2}) is given by
\begin{eqnarray}
g_{k}(x,y) \, & \, = \, & \, a_{+}(x,y)
\eta_{+}^{k}(x) + a_{-}(x,y) \eta_{-}^{k}(x) ,
\label{gk}
\end{eqnarray}
with
\begin{eqnarray}
\eta_{\pm}(x) \, & \, = \, & \, \frac{1}{2 x}
\left( 1 \pm \sqrt{1 - 4 x^{2}} \right)
 \equiv \eta_{\pm} .
\label{gensol}
\end{eqnarray}
The coefficients $a_{\pm}$ are finally  determined from
the equations for $g_{0}$ and $g_{l}$, (\ref{g1}) and (\ref{g3}),
and one obtains
\begin{eqnarray}
g_{l}(x,y) \, & \, = \, & \,
\frac{x y^{2} \left( \eta_{+} - \eta_{-} \right)}
{\eta_{+}^{l} \left( 1 - x y \eta_{-} \right)^{2} -
\eta_{-}^{l} \left( 1 - x y \eta_{+} \right)^{2} } .
\label{glxy}
\end{eqnarray}
This generating function can be viewed as the Green's function of a
particle propagating on a linear chain of length $l$ with hopping
amplitudes at the edge bonds different from the internal ones \cite{balseiro}.

 From Eqs.~(\ref{gl}) and (\ref{genf}) we see that the  contribution
of a dressed skeleton diagram of length $l$ is determined as
\begin{eqnarray}
G_{l}(w) \, & \, = \, & \, \frac{C_{in}(w)}{w} \, \, \,
g_{l}\!\!\left(  \frac{t C_{in}(w)}{w},
\frac{C_{ex}(w)}{C_{in}(w)} \right)  .
\label{plin}
\end{eqnarray}

For the special values of the variables of $g_{l}(x,y)$ appearing in
Eq.~(\ref{plin}) the expression (\ref{glxy}) simplifies considerably
since: $1 - x y \eta_{+}(x) \equiv 0$. We therefore obtain the very
simple final result:
\begin{eqnarray}
G_{l}(w) \, & \, = \, & \, G_{RPA}(w)
\left( \frac{2t}{w + \sqrt{w^{2}-w_{RPA}^{2}}}
\right)^{l}  ,
\label{gesq}
\end{eqnarray}
which holds even for $l = 0$, although our derivation did not include
this case.

In terms of these dressed skeleton diagram contributions we can now write
the propagator as:
\begin{eqnarray}
G_{i,j}^{N}(w) \, & \, = \, & \,
\sum_{l = 0}^{\infty} K_{i,j}^{l} G_{l}(w)  ,
\label{gsn}
\end{eqnarray}
where $K_{i,j}^{l}$ is the number of different bare
background restoring skeleton-paths
of $l$ steps between the end-points $i$ and $j$. On a N\'eel background
only skeleton-paths composed of an even number of steps $l$ will contribute,
as sites $i$ and $j$ must belong to the same sublattice. Trivially,
considering that
 $ K_{i,j}^{0} = \delta_{i,j}$, the RPA is contained in the
propagator of Eq.~(\ref{gsn}).

Eq.~(\ref{gsn}) represents a resummation of the Nagaoka expansion (\ref{ges})
motivated by the merits of the RPA. As we will show in the next section,
it can be used quite efficiently because the number of skeleton-paths
of length $l$ is much smaller than the total number of paths of the same
length. As discussed in the Introduction, the contributions from
non-retraceable paths get less important the higher the dimension
of the lattice. Therefore, we have chosen the two non trivial systems
of lowest
dimension to demonstrate the usefulness of Eq.~(\ref{gsn}): the double
chain and the square lattice.

The resummation (\ref{gsn}) opens a new chance for  a qualified discussion
of the band-edge problem. Brinkman and Rice \cite{brice} have given an
argument which would suggest that the spectrum of the correlated hole
motion on a N\'eel background extends to the Nagaoka energy $w_{FP} = Z t$,
via exponential band-tails. Sorella et al. \cite{sorella} appear to
provide evidence in favour of this scenario by showing that their
spectrum extrapolates to the Nagaoka energy. In our opinion, both the
Brinkman-Rice argument and the Sorella et al. evidence are not
conclusive. It is, in fact, obvious from a total spin decomposition of the
N\'eel state that the spectrum extends to $w_{FP}$ for finite size
systems: the maximum spin $S = L/2$ is contained in a $L$-site N\'eel
background with a probability of order $2^{-L}$. Therefore the issue is
not at all whether the spectrum extends to $w_{FP}$ for systems of
sufficient size, as demonstrated by the extrapolation in \cite{sorella},
but rather whether this part of the spectrum survives with a
finite weight in the thermodynamic limit. The results we present
in the next section point to band-edges below $w_{FP}$.

The position of the band-edge is related to the radius of convergence of
the series in Eq.~(\ref{gsn}). If we replace the series by any
finite summation the band-edge remains at the RPA value (\ref{wrpa}). An
extension of the spectrum beyond this value can only result from the
series diverging for energies $w \leq w_{c}$ with $w_{c} > w_{RPA}$.
One can quite
generally assume an exponential increase in the  number of skeletons
with the path-length:
\begin{eqnarray}
K_{i,j}^{l} \, & \, \propto \, & \,\alpha^{l}, \, \,
\qquad (l \rightarrow \infty).
\label{kalfa}
\end{eqnarray}
This
determines the radius of convergence, which
using Eq.~(\ref{gesq}) gives the following simple relation
between the position of the band-edge $w_{c}$ and the
asymptotic growth parameter, $\alpha$,
of the number of skeleton-paths:
\begin{eqnarray}
w_{c} \, & \, = \, & \, \left\{
\begin{array} {c@{\quad , \quad }l}
w_{RPA} & \alpha \leq \sqrt{Z-1} \\
\left(\alpha + \frac{Z-1}{\alpha}\right) t & \alpha \geq \sqrt{Z-1} \, \, .
\end{array}
\right.
\label{wc}
\end{eqnarray}

The band-edge energy given by (\ref{wc}) grows monotonically with
increasing $\alpha$ until it reaches $w_{FP}$ for $\alpha = Z-1$.
The total number of non-retraceable skeleton-paths of length $l$
for the hole, irrespective
of whether the spin background is restored or not, is $Z(Z - 1)^{l-1}$.
Thus $\alpha = Z-1$ would mean that the number of N\'eel background-restoring
skeleton-paths has the same growth behaviour as the total number of
skeleton-paths. This is extremely implausible at dimensions $D \geq 3$.
In this case the majority of non-retraceable closed random paths
is largely free of
points visited more than once, which means that to restore a N\'eel
background the hole essentially has to walk twice through
such a path. This implies that $\alpha $ should be close to
$\sqrt{Z-1}$, rather than to $Z-1$. The above argument gets better the
higher the dimensionality.

We conclude from the above consideration that we should expect the band-edge
to be close to $w_{RPA}$, that i.e. far away from $w_{FP}$, for lattices
at high dimension. At the same time, we wish to emphasize that we have not
been able to make the above argument exact by a strict estimate of
an upper bound smaller than $Z-1$ for the parameter $\alpha$.
We will see, however, in the next
section that, even for the low dimensional systems we have considered,
our numerical data provide good evidence for the conjecture that
$\alpha < Z-1$ is generally true.

To calculate exactly the propagator given by Eq.~(\ref{gsn}), one would need
to determine the number of distinct background-restoring bare skeleton-paths
of all lengths, which is a very complicated problem.
In the next section we describe how we have overcome this problem for the
concrete cases of a double chain and a square lattice N\'eel backgrounds.

\section{RESULTS AND DISCUSSION}
\label{sec:results}

As mentioned in the previous section, we did find a way to
estimate the numbers of different bare skeleton-paths for all lengths
and thus, through Eq.~(\ref{gsn}), obtain the propagator of a hole
on a N\'eel background in two special cases. In this section,
we will first describe our estimation in general, and in the two
following
subsections present and discuss the results we obtained for the
density of states and spectral density, in the cases of a double chain
and a square lattice.

To be specific, we first determined numerically the exact numbers of different
N\'eel background-restoring skeleton-paths for as many path-lengths
as our computer facilities allowed. For the case of a double chain
we obtained all skeleton-paths up to a length of 32 steps, and
for lengths 34 and 36 we only determined the closed paths. The
results are presented in Table~\ref{table1}. For the
square lattice N\'eel background we obtained all skeleton-paths up to
length 24. In Table~\ref{table2} we exhibit these numbers.

 From the numbers of skeleton-paths exactly obtained, we determined
the asymptotic behaviour as a function of path-length exhibited
by our data. This we could adequately fit by
the following functional form depending on three parameters:
\begin{eqnarray}
K_{i,j}^{l} & \simeq & \frac{C_{i,j} \, \alpha^{l}}{l^{\beta_{i,j}}} ,
\qquad \qquad \forall \, i,j.
\label{fit}
\end{eqnarray}
Notice the exponential dependence with the length which leads to
the band-edge value given by Eq.~(\ref{wc}), as described in the previous
section. We determined the parameter $\alpha$ from fitting
the data for the local propagator, the numbers of closed skeleton-paths. The
band edge of the energy spectrum which we determine
from the local propagator is a ``global" quantity,
in the sense that it has to contain all poles of the
Green's function, and
the non-local Green's functions should not extend beyond this edge.
Taking the latter fact into account
and also to avoid spurious singularities in the self-energy obtained,
we took the parameter $\alpha$  from the local
propagator fit and fixed it while fitting the non-local propagators.
The denominator in Eq.~(\ref{fit}), depending on the parameter $\beta$,
 will essentially account
 for the energy-dependence of the propagator near the band-edge.
Finally we included a proportionality constant, the parameter $C$.

In Tables~\ref{table3} and \ref{table4} we detail the asymptotes obtained
for the double chain and the square lattice respectively.
The asymptotes were obtained by a weighted fit to Eq.~(\ref{fit}) of our
exact short range data,
attaching to the latter a dispersion decreasing with path-length. We
employed a general linear least squares fit solved by singular value
decomposition \cite{numrecipes}.
In particular, for the
numbers of different skeleton-paths of $l$ steps
extending from the origin $i$ to the end-point $j$ defined by the propagator
considered,
which were obtained numerically, we
 considered a standard deviation $\sigma_{l}$:
\begin{eqnarray}
\sigma_{l} \,  & \, \propto \, & \, e^{-0.3 l} .
\label{wfit}
\end{eqnarray}
Considering all data equally  weighted one arrives at asymptotes
similar to those presented here,
but taking a path-length dependent weight
attached to the exact short range data will certainly
produce a better representation of the asymptotic behaviour for
long path-lengths. To quantify this, we can mention that the relative
deviation of the number of skeleton-paths from the asymptotes tabulated
is of the order of $10^{-4}$(double chain) and $10^{-3}$(square lattice)
for the longest paths exactly investigated.
With the weighted fit (\ref{wfit}) which
we employed we could minimize spurious effects in our results, such as
negative peaks in the spectral density. It is important to remark that
the $\alpha$ parameter we obtain, which determines the band-edge,
is very stable to the consideration of different fits. We checked this
by comparing weighted fits (\ref{wfit}) with other exponents of $e^{-l}$,
including the equal-weight case, and
considering other functional forms for the standard deviations
of the data.
In all cases the $\alpha$ parameter
obtained was well within $2\%$ of the value presented in this work for
the double chain, and $8\%$ for the square lattice where fewer exact
data from which to obtain the asymptote are available.

Having obtained the asymptotes for the skeleton-path numbers,
and taking into account that for short path-lengths (say
up to length $l_{0}$) we know
the exact values of these ($K$) numbers, we
evaluated the propagator given by Eq.~(\ref{gsn}) in the following way:
\begin{eqnarray}
G_{i,j}^{N}(w) \, & \, = \, & \, G_{RPA}(w) \delta_{i,j} +
 \sum_{l \geq 2}^{\infty}
\left(\frac{C_{i,j} \alpha^{l}}{l^{\beta_{i,j}}}\right) G_{l}(w)
+ \, \nonumber \\ & \, \, & \,
\sum_{l \geq 2}^{l_{0}} \left( K_{i,j}^{l} -
\frac{C_{i,j} \alpha^{l}}{l^{\beta_{i,j}}} \right) G_{l}(w)  .
\label{gappr}
\end{eqnarray}
The second term in the above expression represents our estimation
of the asymptotic or long skeleton-path contributions
extrapolated to all path-lengths,
while the last term is a correction to include exactly the short range
contributions evaluated numerically.
It is an important detail that the infinite series appearing
in Eq.~(\ref{gappr}) is related to the
polylogarithm function:
\begin{eqnarray}
f_{\beta}(z) \, & \, = \, & \, \sum_{n=1}^{\infty} \frac{z^{n}}{n^{\beta}} ,
\label{poly}
\end{eqnarray}
for which the analytic continuation beyond the radius of convergence
($ | z |  = 1$) is well known.
We obtained the numerical values for this function
employing the Mathematica program \cite{mathem}.

Having evaluated the local propagator as described by Eq.~(\ref{gappr})
we determined the density of states $\rho(w)$:
\begin{eqnarray}
\rho(w) \, & \, = \, & \, - \frac{1}{\pi} Im[ G_{i,i}(w) ] ,
\label{rho}
\end{eqnarray}
and the Fourier transform of the propagator in real space as:
\begin{eqnarray}
G({\bf k},w) \, & \, = \, & \, \sum_{{\bf R}_{j}} e^{- i
{\bf k.R_{\rm j}}} G_{0,j}(w) .
\label{gkw}
\end{eqnarray}
We define the spectral density as:
\begin{eqnarray}
A({\bf k},w) \, & \, = \, & \, - \frac{1}{\pi} Im[ G({\bf k},w) ] ,
\label{akw}
\end{eqnarray}
and the self-energy $\Sigma({\bf k},w)$ by:
\begin{eqnarray}
G({\bf k},w) \, & \, = \, & \, \frac{1}{w - \Sigma({\bf k},w)} .
\label{skw}
\end{eqnarray}
In the following we will present the results we obtained for these
quantities, and discuss the similarities with the results which Sorella
et al. \cite{sorella} obtained
employing the Lanczos spectra decoding method.

\subsection{Double Chain}

In Fig.~\ref{figroch} we plot the density of states obtained for
the double chain with the method described above, and, to compare,
include the RPA density of states in the figure.
We obtain a shift of the band edge, $w_{c} = 2.94 t $, from the
RPA value ($w_{RPA} = 2.83 t$) towards the Nagaoka energy ($w_{FP} = 3 t$),
in accordance with the $\alpha$ parameter obtained (see Table~\ref{table3}):
$\alpha_{RPA} = \sqrt{2} < \alpha \simeq 1.88 < \alpha_{FP} = 2$. This
result was referred to already in Section~\ref{sec:method}.
The good estimation we obtain for $\alpha$ (independent of
the asymptotic fit within $2\%$) and thereby for the band-edge,
indicates the absence of band-tails
extending to the Nagaoka energy for the double chain. The parameter
$\beta$ determines the shape of the density of states near the
band-edges via $ \rho(w) \propto (w_{c}^{2} - w^{2})^{\beta-1} $ .

We obtain a density of states remarkably similar to
Sorella et al. \cite{sorella}, not only in the main features present but
quantitatively as well. In particular, there is a broad peak-like
structure near the band-edge, which we identified as due to the asymptotic
behaviour, second term in Eq.~(\ref{gappr}), corresponding to
long skeleton-path contributions to $G_{i,i}$.
In fact, short range contributions, which are exactly accounted for
by the third term in Eq.~(\ref{gappr}),
are producing a partial cancellation of the
peak which would be obtained due to long skeleton-path contributions.
In the density of states
we can also identify a broad shoulder at intermediate energies
and a second shoulder at lower energies as arising due to the
short skeleton-path contributions, which we exactly calculate here.

To compare the orders
of magnitude of the different contributions involved in $G({\bf k},w)$ of
Eq.~(\ref{gkw}),
we depict in Fig.~\ref{figimg}  the imaginary part of the
local propagator and the most relevant
non-local Green's functions between nearest
neighbours. We recognize a quick decrease in relative weight as one considers
neighbours further apart.
This means that we obtain good convergence for the Fourier transform
of the Green's function, $G({\bf k},w)$, by including the contributions
up to tenth nearest-neighbours.

In Fig.~\ref{figspec} we plot the spectral density (\ref{akw})
as a function of
energy, for three different values of momentum ${\bf k}$. The spectral
weight distribution is in general agreement with the results of Sorella et al.
\cite{sorella}. Concretely, for ${\bf k}
= (0,0)$  we obtain a peak-like structure near the band edge and a broad
shoulder at lower energies, both features present in
Ref.~\cite{sorella} with slight differences
 in magnitude. We investigated the origin of those
features,
and found that the band-edge peak is determined by the asymptotic or
longer skeleton-path contributions to the ${\bf k}$-dependent
Green's function (where
the local and the non-local propagator to $2^{nd}$ nearest neighbours
 are the dominant terms). The short range contributions tend to narrow
and decrease the weight of the peak.
It's interesting to note that
the non-local propagators contributing to $G({\bf k},w)$
tend to shift the peak arising from the local propagator
towards the band-edges and to increase its weight.
 Meanwhile, the broad shoulder at intermediate
energies originates from the short range contributions to the non-local
propagators (mainly those corresponding
to $2^{nd}$ and $3^{rd}$-nearest neighbours).
In our case the main peak appears
accompanied by a narrow secondary peak which is a totally
spurious trace of the RPA band edge, mainly appearing in
the non-local propagators, of no relevance whatsoever.

To exhibit the
dispersion, in Fig.~\ref{figspec} we also plot the
spectral density for ${\bf k} = (\frac{3 \pi}{4},0)$ and ${\bf k} = (\pi,0)$.
Again here we observe the same weight distribution exhibited
by the data of Sorella et al.\cite{sorella}, where $A({\bf k} = (\frac{3 \pi}
{4},0),w)$ contains the highest peak-like structure near the band-edge
in the whole Brillouin zone, while this structure is broadest for the
${\bf k} = (\pi,0)$ data. Also the peak-shifts towards
the center of the band agree with those
obtained in Ref.~\cite{sorella}. For the ${\bf k} \neq 0$ cases,
we find that the short range part of the
non-local propagators contributing to $G({\bf k},w)$
are increasing the weight of the peak which would arise from the local
propagator and shifting it towards the
center of the band.
The same short range part of the non-local propagators
is responsible for the decrease in
spectral weight obtained at intermediate energies (where for
${\bf k} = (0,0)$ the broad shoulder appeared) and for the increase
of weight in the center of the band.

Finally, in Fig.~\ref{figself} we plot the self-energy obtained for the
same ${\bf k}$-values taken for the spectral density results,
and, for purposes of comparison,
we include the dispersionless RPA self-energy.
The effects of
dispersion are clearly observable, and important departures from RPA
are obtained.
In Fig.~\ref{figself}~(a) we plot the real part of the self-energy.
The ${\bf k} = (0,0)$ data show
maximum departures from RPA for energies near the center of the band,
while for the other ${\bf k}$-values these appear shifted towards
intermediate energies.
The origin of these departures lies mainly in the short range parts
of the non-local propagators. Similar comments apply to the imaginary
part of the self-energy plotted in Fig.~\ref{figself}~(b).
For ${\bf k} = (0,0)$,
we obtain an important reduction of the lifetime as compared to the RPA
for energies near the center of the band.

\subsection{Square Lattice}

In Fig.~\ref{figrosq} we plot the density of states we obtained for
the square lattice.
As for the double chain, we obtain a shift of the band-edge, $w_{c} = 3.61 t$,
from the
RPA value ($w_{RPA} = 3.46 t$) towards the Nagaoka energy ($w_{FP} = 4 t$).
This is determined by the $\alpha$ parameter obtained from the
asymptotic fit (see Table~\ref{table4}) through Eq.~(\ref{wc}):
$ \alpha_{RPA} = \sqrt{3} < \alpha \simeq 2.31 < \alpha_{FP} = 4$.
As commented before, the reliability of our estimation
for $\alpha$ (within $8\%$) clearly indicates the absence
of band-tails extending to the Nagaoka energy in the spectrum of
the square lattice. This result also confirms our conjecture of
Section~\ref{sec:method} that the band-edge departures from the
Nagaoka energy increase with dimensionality.

The almost featureless density of states, with minor
departures from the RPA result as expected from the estimations
in terms of the dimension
mentioned in the Introduction, agrees with that obtained by
Sorella et al. \cite{sorella} apart from some minor oscillations.
These oscillations result
from the short skeleton-path contributions to the local propagator.
In contrast to the double chain, here the only observable departure
from the RPA density of states resulting from
the asymptotic behaviour is the band-edge shift.

In Fig.~\ref{figimgsq} we depict the imaginary part of the
local propagator, and the most relevant non-local Green's
functions between nearest-neighbours.
Again, a quick decrease in weight appears on considering
neighbours further apart, but this plot shows that non-local propagators
have much less weight than the local one in comparison to the
double chain (see also Fig.~\ref{figimg}).
This agrees with the estimation of the importance of
non-local corrections to the RPA propagator as a function of the
dimension mentioned in our Introduction. For the square lattice we
obtained the spectral density including in the Fourier transform of
the Green's function, $G({\bf k}, w)$, contributions up to those of twelfth
nearest-neighbours.

In Fig.~\ref{figspecsq} we plot the spectral density as a function of
energy, for three different values of momentum ${\bf k}$. Here for
${\bf k} = (0,0)$  we obtain a low peak-like structure near the band edge
and a shoulder at lower energies, which is broader than the one
in Ref.~\cite{sorella} where the data seem to have been insufficient
to enable them to trace
the complete spectral density curve.
Both structures result mainly from the short range part of
the non-local propagators contributing to $G({\bf k},w)$.
The effect of
dispersion on the spectral weight distribution
is coincident with that shown by Sorella et al. \cite{sorella}.
In Fig.~\ref{figspecsq} we also plot the
spectral density for ${\bf k} = (\pi,0)$,
 which exhibits the highest and broadest peak-like structure,
 and ${\bf k} = (\frac{\pi}{2},\frac{\pi}{2})$, which
shows three maxima, all in accordance with Ref.~\cite{sorella}. Again,
the short range part of
the non-local propagators contributing to $G({\bf k},w)$ accounts for
this distribution of weight.

In Fig.~\ref{figselfsq} we plot the self-energy obtained for the
same ${\bf k}$-values taken for the spectral density, including
the dispersionless RPA self-energy for hole motion on a square lattice
with N\'eel spin order. The qualitative effects of dispersion are similar to
those obtained for the double chain, but the departures from RPA
obtained are smaller, as expected.
As before, the departures obtained mainly result
from the short range part of the non-local propagators.

\section{SUMMARY}
\label{sec:summary}

In this work we studied the motion of a hole on a N\'eel background,
neglecting spin fluctuations, in the framework of the Nagaoka expansion
for the Green's function. Starting from the retraceable-path approximation,
known to become  exact in the one- and the infinite dimensional cases,
where no dispersion appears, we derived a resummation of the
Nagaoka expansion by considering non-retraceable skeleton-paths
and dressing them
by retraceable-path insertions. The contribution of each of such dressed
skeleton-paths was  evaluated exactly. The problem is then reduced
to the determination of the numbers
of different bare skeleton-paths of each length.
We numerically exactly obtained
these numbers up to length 36 for the double chain, and 24 for the square
lattice. From these numbers, we determined an asymptotic extrapolation
for all path-lengths, and used all this information to evaluate the
Green's function. We then determined the density of states and
spectral density for the double chain and square lattice cases.
We determine the band-edges through a
growth parameter, $\alpha$, obtained in
the asymptotic fit. The reliability of our determination of
this parameter, placing the band-edge at a value between the
RPA edge and the Nagaoka energy, points towards the absence
of band-tails extending to the Nagaoka energy in the spectrums of
the double chain and the square lattice.
At the same time, this  confirms our conjecture about the increase of
the band-edge departures from the
Nagaoka energy with dimensionality.
Our results deviate from the exact solution of the problem only
due to the differences between our asymptotic extrapolation and the
unknown exact numbers of skeleton-paths longer than 36 steps for the
double chain and 24 for the square lattice. That these deviations are
small is confirmed by
the general coincidence of our density of states and
spectral density results with those
obtained through the Lanczos approach by Sorella et al. \cite{sorella}.
Our ``analytic" approach has the added advantage of determining the
relevant hole-paths responsible for the main features present in
the density of states and the spectral density.

\acknowledgments

This research was performed within the scientific program of
the Sonderforschungsbereich 341,
supported by the Deutsche Forschungsgemeinschaft.
We would like to acknowledge useful discussions and
comparison with Lanczos results by
C. A. Balseiro and K. A. Hallberg. We thank S. Sorella for sending us
his results prior to publication, and
P. Horsch, D. Vollhardt and J. M. J. van Leeuwen,
 for discussions on their related work.

\newpage

\figure{Double chain. Density of states: our result (full line);
retraceable-path approximation, RPA, (dot-dashed line). \label{figroch}}
\figure{Double chain. $-$ Imaginary part of the local propagator(full line),
of the RPA(local) propagator (dot-dashed line);  of the
non-local propagator between the origin and $2^{nd}$ (dashed line),
$3^{rd}$ (boxes), and $6^{th}$ nearest neighbours (crosses).
 \label{figimg}}
\figure{Double chain. Spectral density $A({\bf k}, w)$ as a
function of energy for: ${\bf k} = (0,0)$ (full line),
${\bf k} = (\frac{3\pi}{4},0)$ (dashed line), ${\bf k} =
(\pi,0)$ (dot-dashed line).
 \label{figspec}}
\figure{Double chain. Self-energy $\Sigma({\bf k},w)$ of the hole
as a function of energy: (a) Real part.
(b) $-$ Imaginary part. RPA (boxes); ${\bf k}=(0,0)$ (full line);
${\bf k}=(\frac{3 \pi}{4},0)$ (dashed line);
${\bf k}=(\pi,0)$ (dot-dashed line).
\label{figself}}
\figure{Square lattice. Density of states: our result (full line);
retraceable-path approximation, RPA, (dot-dashed line). \label{figrosq}}
\figure{Square lattice. $-$ Imaginary
part of the local propagator(full line),
of the RPA(local) propagator (dot-dashed line);  of the
non-local propagator between the origin and $2^{nd}$ (dashed line),
$3^{rd}$ (boxes), and $5^{th}$ nearest neighbours (crosses).
 \label{figimgsq}}
\figure{Square lattice. Spectral density $A({\bf k}, w)$ as a
function of energy for: ${\bf k} = (0,0)$ (full line),
${\bf k} = (\pi,0)$ (dashed line), ${\bf k} = (\frac{\pi}{2},\frac{\pi}{2})$
(dot-dashed line).
 \label{figspecsq}}
\figure{Square lattice. Self-energy $\Sigma({\bf k},w)$ of the hole
as a function of energy: (a) Real part.
(b) $-$ Imaginary part. RPA (boxes); ${\bf k}=(0,0)$ (full line);
${\bf k}=(\pi,0)$ (dashed line);
${\bf k}=(\frac{\pi}{2},\frac{\pi}{2})$ (dot-dashed line).
\label{figselfsq}}

\newpage

\widetext
\begin{table}
\caption{Exact numbers of non-retraceable N\'eel background-restoring
skeleton-paths on a double chain, classified
according to path-length or number of steps (l), and distance (d)
between origin and
end-point, measured in units of the lattice parameter.
Due to symmetry in the double chain,
$n_{d^{2}}^{s}=2$ different sites are end-points
with the same $d (\neq 0)$, thereby
contributing equally to $n_{(d^{2})}^{l}$.}\label{table1}
\begin{tabular}{clccrccccc}
\multicolumn{1}{c}{l}&
\multicolumn{1}{c}{$n_{(d^2=0)}^{l}$} &\multicolumn{1}{c}{$n_{(d^2=2)}^{l}$}
&\multicolumn{1}{c}{$n_{(d^2=4)}^{l}$} &\multicolumn{1}{c}{$n_{(d^2=10)}^{l}$}
&\multicolumn{1}{c}{$n_{(d^2=16)}^{l}$} &\multicolumn{1}{c}{$n_{(d^2=26)}^{l}$}
&\multicolumn{1}{c}{$n_{(d^2=36)}^{l}$}&\multicolumn{1}{c}{$n_{(d^2=50)}^{l}$}
&\multicolumn{1}{c}{$n_{TOTAL}^{l}$}\\
\tableline
6 & 0 & 4 & 0 & 0& 0& 0& 0& 0& 4 \\
10 & 0 & 4 & 8 & 0& 0& 0& 0& 0& 12 \\
12 & 4 & 4 & 6 & 0& 0& 0& 0& 0& 14 \\
14 & 18 & 8 & 12 & 20& 0& 0& 0& 0& 58 \\
16 & 36 & 56 & 30 & 28& 0& 0& 0& 0& 150 \\
18 & 120 & 132 & 66 & 56& 42& 0& 0& 0& 416 \\
20 & 270 & 488 & 302 & 192& 100& 0& 0& 0& 1352 \\
22 & 846 & 1336 & 808 & 364& 254& 96& 0& 0& 3704 \\
24 & 2400 & 3736 & 2610 & 1468& 880& 300& 0& 0& 11394 \\
26 & 7052 & 11180 & 7812 & 4476& 2236& 880& 214& 0& 33850 \\
28 & 21432 & 32500 & 23006 & 14756& 7584& 3376& 844& 0& 103498 \\
30 & 63538 & 99900 & 71162 & 45688& 23898& 10100& 2836& 476& 317598 \\
32 & 193448 & 304628 & 217352 & 141920& 78162& 35228& 11428& 2280& 984446 \\
34 & 590154 & - & - & -& -& -& -& -& - \\
36 & 1824844 & - & - & -& -& -& -& -& - \\

\end{tabular}
\end{table}

\widetext
\begin{table}
\caption{Exact numbers of non-retraceable N\'eel background-restoring
skeleton-paths on a square lattice, classified
according to path-length or number of steps (l), and distance (d)
between origin and
end-point, measured in units of the lattice parameter.
Due to symmetry,
$n_{d^{2}}^{s}$ different sites are end-points with the same $d$ (see
row in brackets).
 }\label{table2}
\begin{tabular}{clccrccccccc}
\multicolumn{1}{c}{l}&
\multicolumn{1}{c}{$n_{(d^2=0)}^{l}$} &\multicolumn{1}{c}{$n_{(d^2=2)}^{l}$}
&\multicolumn{1}{c}{$n_{(d^2=4)}^{l}$} &\multicolumn{1}{c}{$n_{(d^2=8)}^{l}$}
&\multicolumn{1}{c}{$n_{(d^2=10)}^{l}$} &\multicolumn{1}{c}{$n_{(d^2=16)}^{l}$}
&\multicolumn{1}{c}{$n_{(d^2=18)}^{l}$}&\multicolumn{1}{c}{$n_{(d^2=20)}^{l}$}
&\multicolumn{1}{c}{$n_{(d^2=26)}^{l}$}&\multicolumn{1}{c}{$n_{(d^2=32)}^{l}$}
&\multicolumn{1}{c}{$n_{TOTAL}^{l}$}\\
\multicolumn{1}{c}{$(n_{d^{2}}^{s})$}&\multicolumn{1}{c}{(1)}&
\multicolumn{1}{c}{(4)}&\multicolumn{1}{c}{(4)}&
\multicolumn{1}{c}{(4)}&\multicolumn{1}{c}{(8)}&
\multicolumn{1}{c}{(4)}&\multicolumn{1}{c}{(4)}&
\multicolumn{1}{c}{(8)}&\multicolumn{1}{c}{(8)}&\multicolumn{1}{c}{(4)}
& \\
\tableline
6 & 0 & 8 & 0 & 0& 0& 0& 0& 0&0&0& 8 \\
10 & 0 & 16 & 32 & 0& 0& 0& 0& 0& 0&0&48 \\
12 & 8 & 16 & 24 & 32& 0& 0& 0& 0& 0&0&80 \\
14 & 72 & 192 & 120 & 32& 80& 0& 0& 0&0&0& 496 \\
16 & 440 & 528 & 384 & 208& 368& 0& 0& 0& 0&0&1928 \\
18 & 1728 & 2912 & 1488 & 768& 1056& 168&0& 128& 0&0&8248 \\
20 & 8512 & 12176 & 7408 & 4072& 5296& 656& 256& 640& 0&0&39016 \\
22 & 33224 & 58648 & 35400 & 15072& 19904& 3224& 2312& 2832& 384&0&171000\\
24 & 151224 &257472 & 159376 & 76712& 102528& 16208& 10608& 13600&
3824&512&792064 \\

\end{tabular}
\end{table}

\mediumtext
\begin{table}
\caption{Double chain:
Asymptotes obtained for $K_{i,j}^{l}$ (as defined by Eq.~(\ref{fit})).
In the first column we enter the relative position of end-point $j$ with
respect to the origin $i$, denoting by $n^{th}(i)$ a $n^{th}$
nearest-neighbour site of the origin $i$. For clarity, in the second column
we write the distance $d$ between $i$ and $j$ (squared).
The last column indicates which
 $(n_{d^{2}}^{l} / n_{d^{2}}^{s})$ data (from Table~\ref{table1})
were employed to determine each
asymptote by the weighted fit (Eq.~(\ref{wfit})), by stating the
respective $l$.}\label{table3}
\begin{tabular}{cccccc}
\multicolumn{1}{c}{$j$}& \multicolumn{1}{c}{$d^{2}$}&
\multicolumn{1}{c}{$\alpha$} &\multicolumn{1}{c}{$\beta_{i,j}$}
&\multicolumn{1}{c}{$C_{i,j}$} &\multicolumn{1}{c}{$l$} \\
\tableline
$j=i$ &0 & 1.88136 & 2.39084 & 1.26191 & 22-36 \\
$2^{nd}(i)$&2 & " & 2.27482 & 0.66616 & 22-32 \\
$3^{rd}(i)$&4 & " & 2.17748 & 0.33958 & 22-32 \\
$6^{th}(i)$ &10& " & 1.76242 & 5.278$\times 10^{-2}$ & 22-32 \\
$7^{th}(i)$ &16& " & 1.42770 & 9.05$\times 10^{-3}$ & 22-32 \\
$10^{th}(i)$ &26& " & 0.86016 & 5.7$\times 10^{-4}$ & 22-32 \\

\end{tabular}
\end{table}

\mediumtext
\begin{table}
\caption{Square lattice:
Asymptotes obtained for $K_{i,j}^{l}$ (as defined by Eq.~(\ref{fit})).
In the first column we enter the relative position of end-point $j$ with
respect to the origin $i$, denoting by $n^{th}(i)$ a $n^{th}$ nearest-
neighbour site of the origin $i$. For clarity, in the second column
we write the distance $d$ between $i$ and $j$ (squared).
The last column indicates which
 $(n_{d^{2}}^{l} / n_{d^{2}}^{s})$ data (from Table~\ref{table2})
were employed to determine each
asymptote by the weighted fit (Eq.~(\ref{wfit})), by stating the
respective $l$.}\label{table4}
\begin{tabular}{cccccc}
\multicolumn{1}{c}{$j$}&\multicolumn{1}{c}{$d^{2}$}&
\multicolumn{1}{c}{$\alpha$} &\multicolumn{1}{c}{$\beta_{i,j}$}
&\multicolumn{1}{c}{$C_{i,j}$} &\multicolumn{1}{c}{$l$} \\
\tableline
$j=i$ &0& 2.31204 & 2.18809 & 0.29007 & 16-24 \\
$2^{nd}(i)$ &2& " & 1.76948 & 3.298$\times 10^{-2}$ & 16-24 \\
$3^{rd}(i)$ &4& " & 1.56262 & 1.056$\times 10^{-2}$ & 16-24 \\
$5^{th}(i)$ &8& " & 1.67271 & 7.07$\times 10^{-3}$ & 16-24 \\
$7^{th}(i)$ &10& " & 1.74651 & 5.95$\times 10^{-3}$ & 16-24 \\
$9^{th}(i)$ &16& " & 1.05138 & 2.1$\times 10^{-4}$ & 18-24 \\
$11^{th}(i)$ &18& " & 1.75567 & 1.29$\times 10^{-3}$ & 22-24 \\
$12^{th}(i)$ &20& " & 1.40003 & 2.7$\times 10^{-4}$ & 18-24 \\

\end{tabular}
\end{table}

\end{document}